\documentclass{ws-procs961x669}            % default, citations in superscript
\usepackage{siunitx}
\begin{document}
\title{Evidence for dynamical changes in Betelgeuse using multi-wavelength data}
\author{Sneha Kachhara}

\address{Indian Institute of Science Education and Research (IISER) Tirupati,\\
Tirupati, India\\snehakachhara@students.iisertirupati.ac.in}

\author{Sandip V. George}

\address{\quad Department of Computer Science, University College London, London WC1E 6BT, United Kingdom\\
\quad Interdisciplinary Center for Psychopathology and Emotion Regulation, \\ University Medical Center Groningen\\ University of Groningen, The Netherlands\\sandip.george@ucl.ac.uk}

\author{Ranjeev Misra}

\address{Inter University Center for Astronomy and Astrophysics (IUCAA) Pune,\\
Pune, India\\rmisra@iucaa.ac.in}
\author{G Ambika}

\address{Indian Institute of Science Education and Research (IISER) Tirupati,\\
Tirupati, India\\g.ambika@iisertirupati.ac.in}

\begin{abstract}

The reasons behind the Great Dimming and subsequent rising in the brightness of Betelgeuse between October 2019 and March 2020 still continues to baffle astronomers. It has been shown by George et. al. (2020) that critical slowing down preceded the dimming event. This suggested that the dimming was as a result of the change in the nature of the nonlinear dynamics of the star. In this work we present additional evidence for dynamical changes in Betelgeuse prior to the Great Dimming event, using nonlinear time series analysis. We study the relations between the different bands in the photometry data collected from the Wing photometery (IR/near-IR) and Wasatonic observatory (V-band). We also analyse how the early warning signals studied previously changed during and after the Great Dimming.
\end{abstract}

\keywords{critical slowing down, alpha orionis, recurrence quantification analysis, wing photometry.}

\bodymatter

\section{Overview}
The unprecedented dimming and subsequent brightening events in Betelgeuse between October 2019 and March 2020 have since been studied in extensive detail\cite{guinan2019fainting, guinan2019updates,sigismondi2020rapid}. Three main hypotheses have been suggested as reasons for the dimming, namely changes in pulsation dynamics, star spots and a dust cloud. Combinations of these hypotheses have also been suggested as reasons for the dimming\cite{dupree2020spatially, kravchenko2021atmosphere}. We will explore each of these explanations below, before explaining how the nonlinear dynamics of the light curve can contribute towards understanding the dimming.

The most prominent reason suggested for the Great Dimming observed in Betelgeuse has been the formation of a dust cloud. This has been put forward as an explanation by multiple authors. Measurements of effective temperature have suggested that no significant drop was observed during the dimming\cite{levesque2020betelgeuse}. In addition, high angular resolution images showed a significant dimming in the southern hemisphere of Betelgeuse during the dimming event\cite{montarges2021dusty}. Observations from multiple wavelengths suggest that an outflow from the star itself, enhanced by pulsations, condensed around the southern hemisphere of the star\cite{dupree2020spatially, kravchenko2021atmosphere}. It has also been suggested that the dimming could be explained by an ejection from the star that cools below 3000 K to form molecules, before condensing to dust\cite{davies2021impact}.

The dust hypothesis has been disputed as the sole reason for the dimming by a number of authors. Radiative transfer models using data from sub-millimeter wavelengths suggested that the dimming was a result of changes in the photosphere as opposed to external dust\cite{dharmawardena2020betelgeuse}. Further, observations during the dimming suggested that there was little change in the IR flux\cite{gehrz2020betelgeuse}. Moreover, Wing three-filter TiO and near-IR photometry found that the effective temperature during the dimming could be much lower than previously calculated, suggesting that the dimming observed in the visible spectrum could be due to photospheric motions\cite{harper2020photospheric}. Many of these authors proposed the presence of spots on the stellar surface as an alternative reason for the dimming of Betelgeuse \cite{dharmawardena2020betelgeuse, alexeeva2021spectroscopic}.

Finally, it has been suggested that the observed dimming was as a result of changes in pulsational dynamics\cite{guinan2020continued, harper2020photospheric, george2020early}.  Multiple authors have pointed out that the Great Dimming coincided with the minima of both the 430 days and 5.8 year periods\cite{harper2020photospheric, percy2020s, sigismondi2020betelgeuse}. However, the observed decrease in temperature in the star could not be explained by pulsational dynamics alone\cite{alexeeva2021spectroscopic}. 

Nonlinear time series analysis on the light curve of Betelgeuse prior to the major dimming event suggested that a critical slowing down was observed in the light curve well before the actual dimming occurred. When a nonlinear dynamical system undergoes a change of state, it often does so suddenly and with no change in the mean response. However, prior to many such changes in state (called a critical transition), multiple time series quantifiers called early warning signals (EWS) are shown to increase\cite{scheffer2009early}. The presence of critical slowing down in Betelgeuse indicated that the star was about to undergo a change of state in its dynamical properties. It was suggested that the dimming phenomenon was the change of state itself or occurred as a result of it\cite{george2020early}. This work suggested that the reasons behind the dimming could be explained by studying the light curve before the Great Dimming occurred.

In this paper we explore the dynamical properties of Betelgeuse in greater detail, by studying how they changed before, during and after the Great Dimming. We do this by analysing Wing three-filter TiO and V band photometry measured at the Wasatonic Observatory\cite{harper2020photospheric,harper2021sofia} and V/Vis data from the American Association of Variable Star Observers (AAVSO)\cite{Kafka2021}. We first gain further evidence for a dynamical change in Betelgeuse by studying the multiwavelength Wing data. We then study changes in the linear correlations between the wavelengths and study how they changed over the period leading up to the dimming. Finally we study how the quantifiers studied by George et. al \cite{george2020early} changed in the AAVSO light curve, in the period leading up to, during and after the Great Dimming.

\section{Analysis}
\subsection{Wing Three-filter and V-band Photometry}
\begin{figure}
    \centering
    \includegraphics[width=\textwidth]{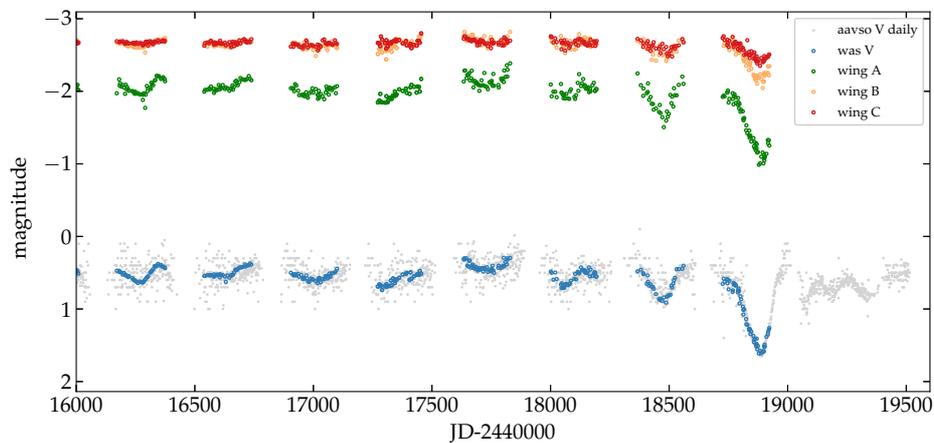}
    \caption{Variation of intensity in Betelgeuse, as captured in Vis/V band AAVSO data (grey), V-band photometry from Wasatonic laboratory (blue), and Infrared Wing three-filter (A: green, B: orange, C: red) data.}
    \label{fig:full_lc}
\end{figure}
%intro to the data
Wing photometry consists of an eight band measurement of electromagnetic radiation from an astronomical source, initially developed for studying M supergiant stars\cite{white1978photoelectric}. Later, a simplified three-filter system was proposed that measures infrared magnitude, color and spectral type for a star\cite{wing1992three}. In this study, photometric data (IR/near-IR) from three Wing filters\cite{wing1992three}, and a wideband V filter collected by the Wasatonic observatory, are analyzed. The V-band filer is centered around 5550 $\AA$. The Wing A filter is dominated by the TiO 7190 $\AA$. The Wing B and C filters are continuum bands centered around 7500 $\AA$ and 10240 $\AA$ respectively. More information about the data may be found in Harper et. al.(2020) \cite{harper2020photospheric}.

\begin{figure}
    \centering
    \includegraphics[width=0.8\textwidth]{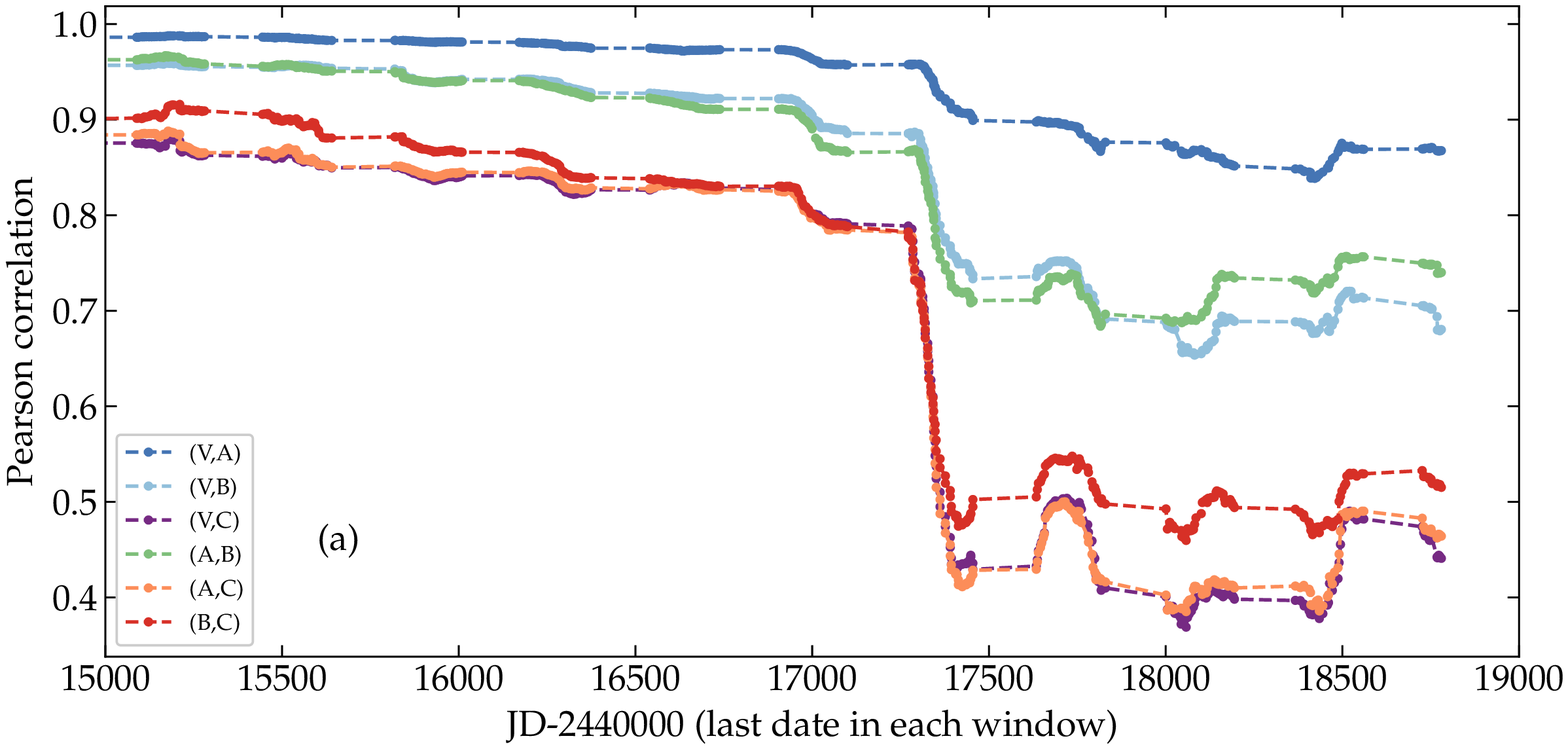}
    \includegraphics[width=0.8\textwidth]{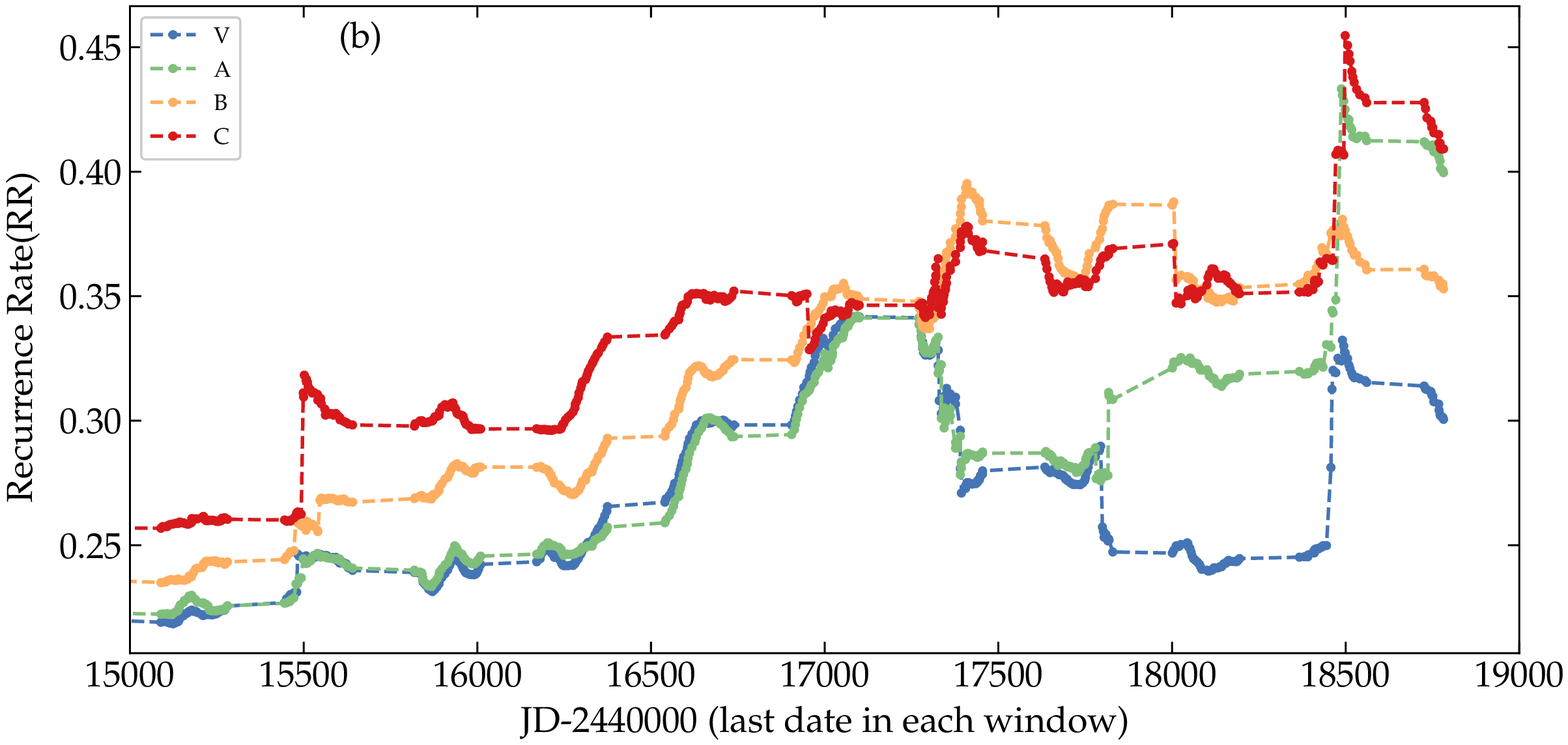}
    \caption{(a) Cross-correlation between pairs of data from different bands calculated in a sliding window fashion upto (but not including) dimming.
    (b) Recurrence Rate (RR) obtained with fixed recurrence threshold = 0.1, for each band. Window size = 300.}
    \label{fig:multi_corr}
\end{figure}

We first calculate the cross-correlations (see figure \ref{fig:multi_corr} (a)), using Person Correlation Coefficient among data from different bands in a sliding window fashion with a window size = 300 points. The V-band and A band are already known to exhibit strong correlation\cite{harper2020photospheric}, but we find that this correlation decreases prior to dimming. It is interesting that this trend starts much before the actual dimming episode, hinting towards a prolonged process. Moreover, we find this trend to be simultaneous in all pairs of data.\\

We then employ nonlinear time series analysis techniques to analyze the multi-variate data. Building on dynamical systems theory, our techniques assume the existence of a state space representation for the data. Since dynamical features of the source are intimately related to the topological properties in state space, its quantification can yield valuable information.%maybe also some details on autocorrelation etc.

One of these helpful properties is recurrence: the tendency of a system to visit the neighborhoods of previously occupied states. Recurrence is a prominent feature of deterministic dynamics and can be easily captured in terms of Euclidean distances among points on the attractor (trajectories) in state space\cite{marwan2007recurrence}. When presented in matrix form, it is known as the Recurrence matrix ($R$). A 2-dimensional visual representation of $R$ is known as the recurrence plot (RP). Statistical measures on RP, collectively known as the Recurrence Quantification Analysis (RQA), reveal subtle characteristics of the dynamics at play. Changes in some of these RQA measures may serve as EWS for a system close to a dynamical transition\cite{marwan2013recurrence}. For the multivariate data, we make use of Recurrence Rate (RR) to study dynamical features of the Betelgeuse light curve.\\

We analyze recurrence properties of the data in the same sliding window fashion, keeping distance threshold constant at 0.1 for calculation of RP. We do this separately for each band, and present the results for RR (fraction of points in the recurrence matrix) in figure \ref{fig:multi_corr} (b). Again, we find an increasing trend in all wavelengths. Interestingly, the trend is most consistent for IR bands. Further, it corresponds well to the correlation indicated in figure \ref{fig:multi_corr} (a). This indicates that if there is a dynamical process at play, it may not be a peripheral phenomenon. Similar variation across bands hints at the possibility of change at the stellar scale, probably in the pulsation dynamics.\\

These results agree well with EWS observed in the AAVSO data\cite{Kafka2021}, reported previously\cite{george2020early}. However, the length of available data limits an extensive analysis, at this point in time. 

\subsection{Beyond the Great Dimming}
In this section we present results on how the nonlinear parameters of the light curve changed after the Great Dimming, using V band data from the AAVSO\cite{Kafka2021}. In George et. al. (2020) \cite{george2020early}, we reported signatures of dynamical transitions in this data in the form of critical slowing down. We reported an increase in multiple EWS including the autocorrelation at lag 1 (ACF(1)), variance and recurrence based measures in the data leading up to the dimming. The increase in these quantifiers indicated that a critical transition was imminent. It was assumed that the Great Dimming occurred as a result of this critical transition. In the present paper we go beyond the Great Dimming and analyse how the EWS changed since. 
\subsubsection{Autocorrelation and Variance}
\begin{figure}[t]
\centering
\includegraphics[width=.45\textwidth]{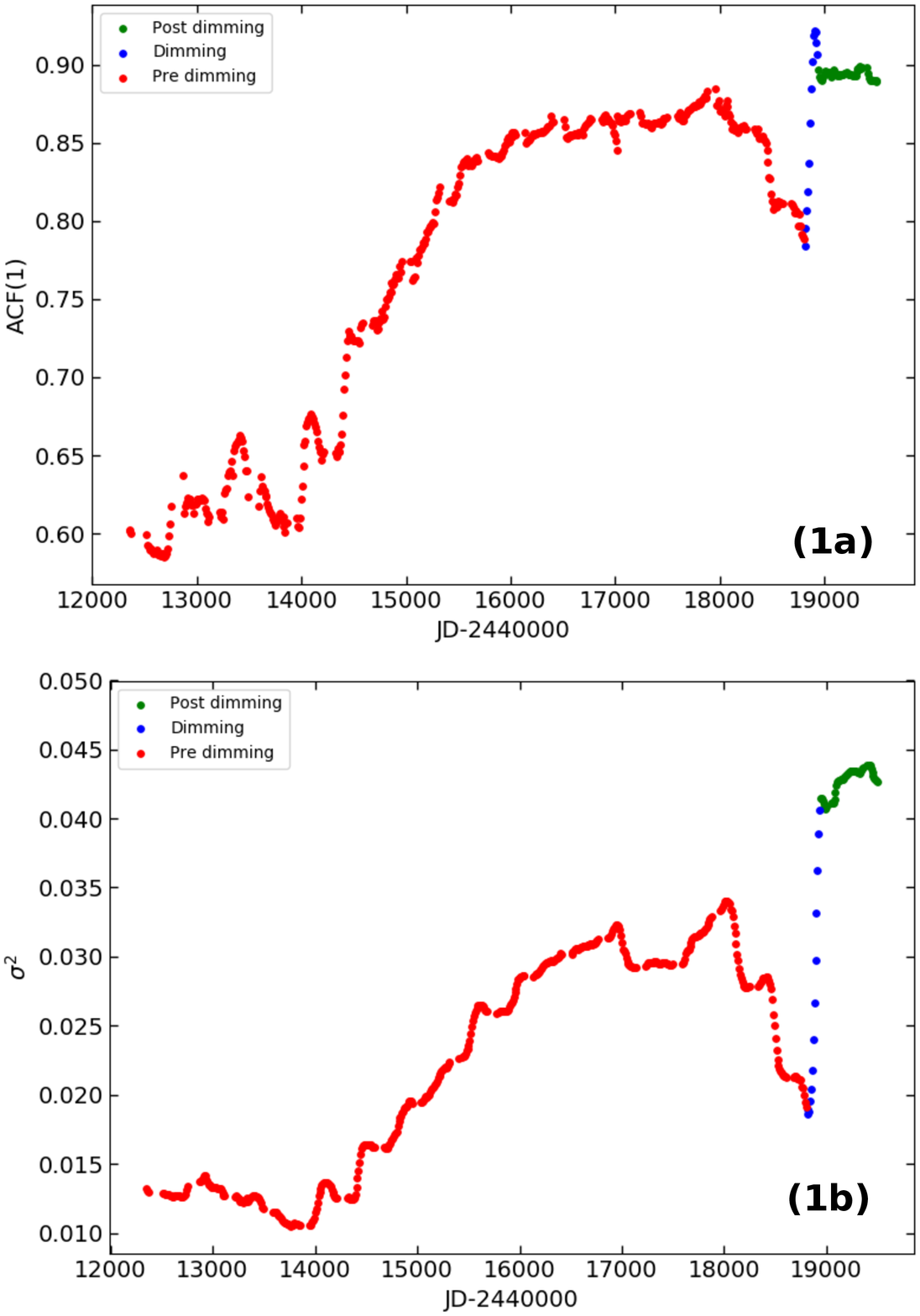}
\includegraphics[width=.45\textwidth]{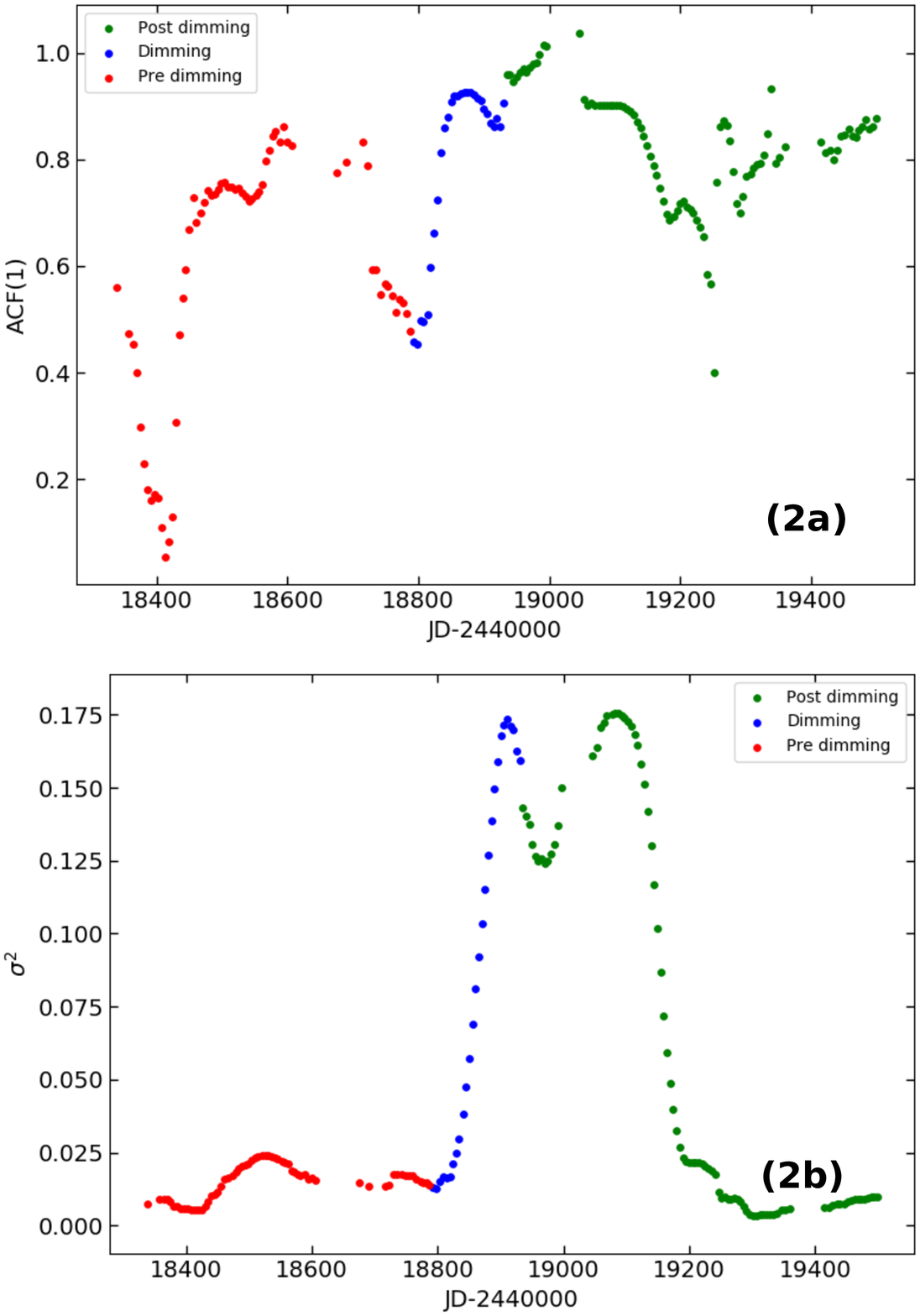}
\caption{Variation in the (a) Autocorrelation at lag 1 and (b) Variance calculated from the light curve of Betelgeuse upto the dimming (red), during the dimming (blue) and post dimming (green). The left panels (figures 1a and 1b) show the variation from 1990 onwards using a window size of 300 points on data binned every 10 days. The right panels zoom into a region around the dimming, with data binned every 5 days with a window size of 40. }
\label{fig:acfvar}
\end{figure}

To do this we initially bin the data from 1990 onwards into 10 day bins, and calculate the ACF(1) and variance in windows of size 300. This is presented in figure \ref{fig:acfvar}.1. Both the ACF(1) and Variance show an increase leading up to the dimming, a sharp rise when the Great Dimming occurred, and a slow reduction after the dimming ended. In figure \ref{fig:acfvar}.2 we select an area around the dimming, bin the data into 5 day bins and choose a window size of 40 points. The data again shows the same trends, with a rise leading up to the dimming, a sharp rise during the Great Dimming and a fall since. With the smaller window and bin size, the error on the quantifiers increases (changing approximately as $\frac{1}{\sqrt{n}}$).

\subsubsection{Recurrence Measures}
\begin{figure}[t]
\centering
\includegraphics[width=\textwidth]{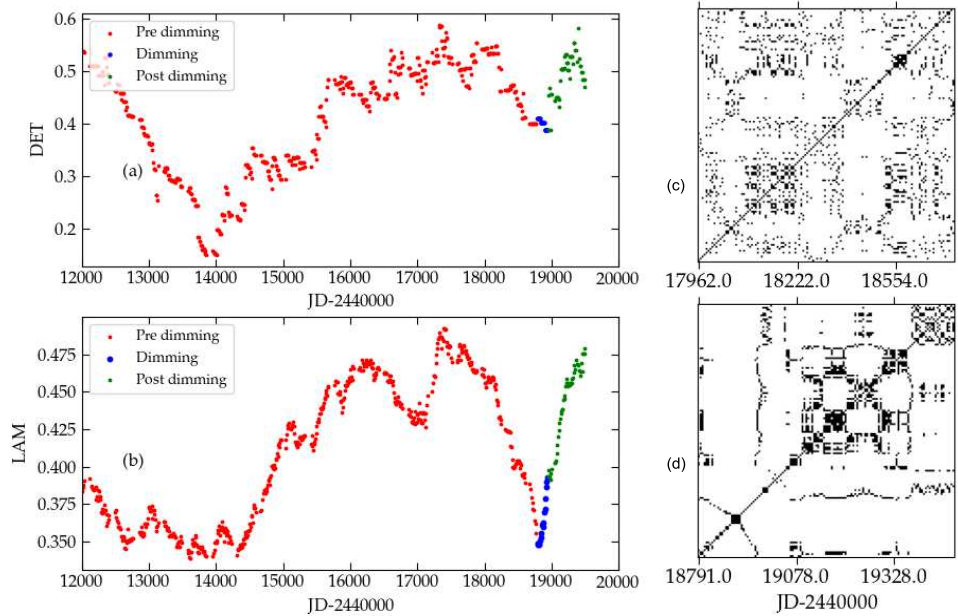}
\caption{Recurrence measures (a) Determinism (DET) and (b) Laminarity (LAM) calculated from the light curve of Betelgeuse upto the dimming (red), during (blue), and post dimming (green). The left panels (figures a and b) show the variation from 1990 onwards using a window size of 300 points on data binned every 10 days. The right panels (figures c and d) show typical recurrence plots for prior and during dimming respectively. The recurrence rate was fixed at 0.1.}
\label{fig:det_lam_aavso}
\end{figure}

We then analyze AAVSO data in terms of recurrence measures as well. For this, we fix the recurrence rate (RR) as 0.1, and calculate Determinism (DET) and Laminarity (LAM) in the same sliding window fashion. We find a consistent rise in these measures prior to dimming, followed by a sudden dip and subsequent rise (see figure \ref{fig:det_lam_aavso} (a) and (b)). We also present typical recurrence plots for two windows near dimming episode (one prior and one including dimming) in figure \ref{fig:det_lam_aavso} (c) and (d) respectively as a visual aid. Clearly, the sudden drop in brightness affects the structure of the RP, and the recurrence measures reflect it. However, it is interesting to see the rise prior to dimming, possibly caused by dynamical factors (details discussed in George et al.\cite{george2020early}). The trend in post dimming behaviour seems in agreement which could be verified with future observations in the coming years.\\

\section{Discussion}
%tying it all together
We presented the nonlinear analysis on two datasets measuring the brightness of Betelgeuse before, during and after the Great Dimming of 2019-20. We conducted a multivariate time series analysis of the data from Wing-IR photometry and the V band photometry from Wasatonic laboratory. We show that prior to the dimming, the cross correlations and individual recurrence measures vary considerably. Our results suggest that a change in the dynamics of Betelgeuse commenced in advance of the Great Dimming of 2019-20. We also analyse early warning signals in AAVSO light curve for the period including and after the Great Dimming. While not conclusive, we see evidence that the values of the early warning signals have reduced since the Great Dimming.

Our results make a strong case that the dynamics of Betelgeuse underwent significant changes leading up to the Great Dimming. Variable stars are known to undergo changes in their nonlinear properties as they evolve\cite{buchler1987period, szabo2010does, george2020fractal}. In this study we observe that the recurrence properties, cross correlations and early warning signals change prior to the Great Dimming. In addition, Betelgeuse is known to exhibit multiple frequencies in its light curve, including a short period, thought to be driven by the $\kappa$ mechanism, a long period thought to be convection driven\cite{stothers2010giant} and a recently detected overtone of the short period\cite{harper2020photospheric, joyce2020standing}. It is possible that the pulsational dynamics of Betelgeuse, including the birth of new periods may have preceded the Great Dimming, and these need to be examined in detail. 

Taken together with the study on the light curve of Betelgeuse by George et. al. (2020)\cite{george2020early}, our results seem to indicate that the Great Dimming was not an isolated incident, and to understand the reasons for the dimming, we must consider changes in the dynamics before it. 
\section*{Acknowledgments}
We acknowledge Dr. Graham Harper for useful discussions, and  Richard Wasatonic and Prof. Edward Guinan for the multi wavelength data used in this study. We acknowledge with thanks the variable star observations from the AAVSO International Database contributed by observers worldwide and used in this research. SVG acknowledges financial support from the European Research Council (ERC) under the European Union’s Horizon 2020 research and innovative programme (ERC-CoG-2015; No 681466 awarded to M. Wichers). SK acknowledges financial support from the Council of Scientific and Industrial Research (CSIR), India.

\end{document}